\PassOptionsToPackage{unicode}{hyperref}
\PassOptionsToPackage{hyphens}{url}
\documentclass[
  11pt,
]{article}
\usepackage{amsmath,amssymb}
\usepackage{iftex}
\ifPDFTeX
  \usepackage[T1]{fontenc}
  \usepackage[utf8]{inputenc}
  \usepackage{textcomp} % provide euro and other symbols
\else % if luatex or xetex
  \usepackage{unicode-math} % this also loads fontspec
  \defaultfontfeatures{Scale=MatchLowercase}
  \defaultfontfeatures[\rmfamily]{Ligatures=TeX,Scale=1}
\fi
\ifPDFTeX\else
\fi
\IfFileExists{upquote.sty}{\usepackage{upquote}}{}
\IfFileExists{microtype.sty}{% use microtype if available
  \usepackage[]{microtype}
  \UseMicrotypeSet[protrusion]{basicmath} % disable protrusion for tt fonts
}{}
\usepackage{xcolor}
\usepackage{longtable,booktabs,array}
\usepackage{calc} % for calculating minipage widths
\usepackage{etoolbox}
\makeatletter
\patchcmd\longtable{\par}{\if@noskipsec\mbox{}\fi\par}{}{}
\makeatother
\IfFileExists{footnotehyper.sty}{\usepackage{footnotehyper}}{\usepackage{footnote}}
\makesavenoteenv{longtable}
\usepackage{graphicx}
\makeatother
\usepackage{pos}
\usepackage[absolute,overlay]{textpos}
\usepackage{tikz}
\usetikzlibrary{math}

\author*[a,b]{Sofie Martins}
\author[a]{Erik Kjellgren}
\author[a]{Emiliano Molinaro}
\author[a,b]{Claudio Pica} 
\author[a,b]{Antonio~Rago}

\affiliation[a]{University of Southern Denmark, Campusvej 55, 5230 Odense M, Denmark}
\affiliation[b]{$\hbar$QTC, University of Southern Denmark, Campusvej 55, 5230 Odense M, Denmark}

\emailAdd{martinss@imada.sdu.dk}

\abstract{HiRep allows flexible simulations of higher representations of Wilson Fermions with various actions and gauge groups and a range of inverters and integrators. This is particularly important for enabling evaluations of observables relevant to phenomenological inputs for Beyond-the-Standard-Model physics from lattice field theory. We present progress on the GPU porting of available features, especially in terms of scaling to large jobs on AMD GPUs.}

\FullConference{The 41st International Symposium on Lattice Field Theory (LATTICE2024)\\
 28 July - 3 August 2024\\
Liverpool, UK\\}

\usepackage{flafter}
\usepackage{float}
\usepackage{pos}
\usepackage{natbib}
\usepackage{subfigure}
\usepackage{booktabs}
\usepackage{tikz}
\usepackage{ifthen}
\usepackage{longtable}
\usepackage{array}
\usepackage{multirow}
\usepackage{wrapfig}
\usepackage{colortbl}
\usepackage{pdflscape}
\usepackage{tabu}
\usepackage{threeparttable}
\usepackage{threeparttablex}
\usepackage[normalem]{ulem}
\usepackage{makecell}
\usepackage{xcolor}
\usepackage[]{natbib}
\nocite{*}
\usepackage{bookmark}
\IfFileExists{xurl.sty}{\usepackage{xurl}}{} % add URL line breaks if available
\hypersetup{
  pdftitle={Scaling SU(2) to 1000 GPUs using HiRep},
  hidelinks,
  pdfcreator={LaTeX via pandoc}}

\title{Scaling SU(2) to 1000 GPUs using HiRep}
\date{\vspace{-2.5em}}

\begin{document}
\maketitle

{
\setcounter{tocdepth}{2}
\tableofcontents
}
\section{Motivation}\label{motivation}

Modern supercomputers reach unprecedented peak bandwidths and computational throughput using Graphics Processing Units (GPUs). Exploiting these peak performances to simulate BSM physics allows the generation of high-precision, non-perturbative predictions for experimental input. While there is support for models with higher representations from numbers of colors of 2 to 5 in Grid \citep{Boyle:2016lbp, Bennett:2023gbe} using Staggered or Wilson Fermions, HiRep \citep{DelDebbio:2008zf, Martins:2024dew} is currently the highest performing option supporting Wilson Fermions with arbitrary numbers of flavors and colors in SU(N) gauge groups with fermions in the fundamental and selected higher representations for CPUs and GPUs. We aim to minimize the time needed for state-of-the-art lattice calculations with unprecedented statistics, supporting possibly full machine runs without loss of efficiency. While the GPU version was developed for NVIDIA GPUs, we reached reasonable efficiency also on AMD machines, such as LUMI-G.

The software is available at

{\vspace{0.5cm}
\centering

\texttt{https://github.com/claudiopica/HiRep} \flushleft
}

\section{Execution structure}\label{execution-structure}

\subsection{Wilson-Dirac Operator}\label{wilson-dirac-operator}

Implementing the Wilson-Dirac operator for multi-GPU runs depends on the execution of different kernels. First, a kernel calculates the inner computations, that is, the computations that do not depend on communicated information from other GPUs. Then, there is a boundary calculation that computes all terms only after communicated information is received. The communications rely on executing a kernel that synchronizes data from the local lattice to a send buffer, followed by a memory transfer.

The efficiency of this kernel depends on the organization of these different kernels and memory operations. They can be analyzed by using profiler tracing data, either from NVIDIA Nsight (CUDA) or \texttt{rocprof} (ROCm). Below, we show traces for LUMI-Gs AMD MI250X GCDs and UCloud DeiC Interactive HPCs NVIDIA H100 GPUs.

In these figures, times are given relative to an arbitrarily chosen starting point chosen as a representative example of the sequence of kernel executions. There is inevitable variability from one Dirac execution to another. The example we show is typical and not the best one we found in the tracing data. Any timing reported here is up to the precision of the profiler's sampling.

Our ability to achieve the peak capabilities of the device is measured in terms of memory bandwidth instead of computational throughput, as lattice simulations are memory-bound.

\begin{figure}

{\centering \includegraphics{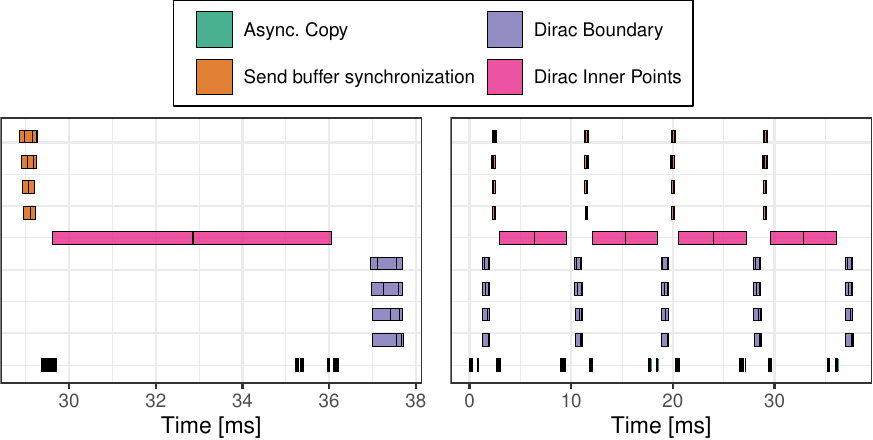} 

}

\caption{ROCm profiles for single and multiple Dirac operator executions on LUMI-G in a 4 GCD/single-node job on a $48^4$ local lattice parallelized in two dimensions. The vertical direction is arranged first by kernel type and then by queue ID, showing the parallelization of different kernel types in different streams. In this case, the profile on the right shows only one of the four MPI processes.}\label{fig:amd-dirac}
\end{figure}

In Fig.~\ref{fig:amd-dirac}, we mainly want to see that the send buffer synchronizations and the Dirac boundary execution are sufficiently parallelized to minimize the overhead. Further, the communications have to be in parallel to the inner Dirac application, and they mainly are. In principle, the split up even and odd parts of the Dirac inner point computations can also be streamed in parallel; in practice, this will not work because the inner Dirac operator kernel uses the GPU to its full extent. Additionally, this parallelization is impossible in the even-odd preconditioned version of the Dirac operator used mainly in the HMC, so this improvement would be of little practical relevance.

\begin{figure}

{\centering \includegraphics{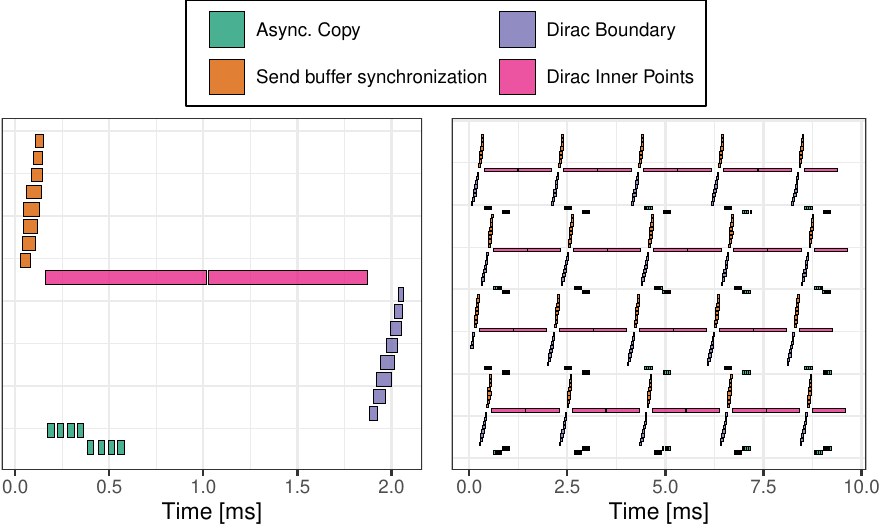} 

}

\caption{Profile of repeated Dirac operator applications on four NVIDIA H100 GPUs using a $48^4$ local lattice parallelized in two dimensions. The tracing plot shows on the left side a single application of the Dirac operator as part of this profile and on the right side the repeated Dirac executions on all four MPI processes. The vertical direction is arranged first by process, then kernel type, and finally by stream ID. In contrast to the ROCm profiles, this shows both the MPI parallelization at the node level and the software-level streaming structure. The ROCm profiles only show a single MPI process and the queue ID at the hardware level.}\label{fig:nvidia-dirac}
\end{figure}

For comparison, we show in Fig.~\ref{fig:nvidia-dirac} the execution structure of a 4-GPU NVIDIA H100 node. The structure in execution is similar, but the node achieves slightly better results in executing the communications in parallel to the heavy inner Dirac kernel, and the overall execution times are significantly reduced, mainly due to the higher theoretical peak bandwidth. Note, that the two profiles are different in displaying the parallelism. While the NVIDIA Nsight trace contains information about all processes and the stream to which the software submitted the kernel, the ROCm traces only show a single process and the hardware level queue that the kernel ends up being scheduled to, regardless of which stream the kernel was selected for.

\subsubsection{Blocking and non-blocking communications}\label{blocking-and-non-blocking-communications}

All communications in \texttt{HiRep} are executed from a different POSIX thread. This means we can either call blocking or non-blocking communications in parallel to the inner Dirac execution. While non-blocking communications have the advantage that they do not pile up requests, sending all requests at once can enable a well-configured network to manage the requests better. Which setup is more efficient needs to be tested on an individual basis.

\subsection{Clover improvement}\label{clover-improvement}

Clover improvement of the Wilson-Dirac operator requires the application of another diagonal term. While the parallelization of this term is trivial, the site-by-site operations require more memory and computation. Therefore, the applications were optimized by precomputing certain fields, such as the LDL-factorization and exponential of the clover term, respectively, in such a way that they need only to be recomputed whenever the gauge field is updated.

\begin{figure}

{\centering \includegraphics{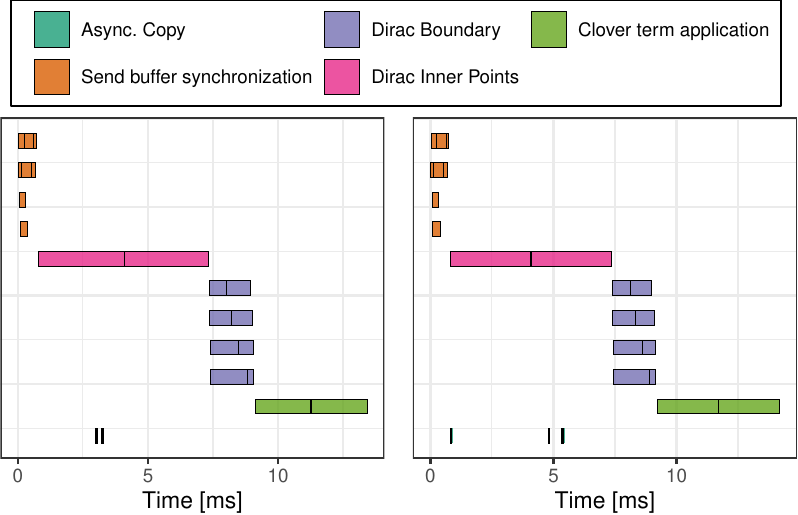} 

}

\caption{A single Dirac operator application including the diagonal clover term executed on AMD MI250X GCDs on LUMI-G using a $48^4$ local lattice with two parallelized dimensions, both for regular clover improvement (left) and exponential clover improvement (right). The vertical direction is organized first by kernel type and second by queue ID.}\label{fig:clover-trace}
\end{figure}

The results in Fig.~\ref{fig:clover-trace} show that the application of the additional clover term only takes little additional time. The extra time needed is approximately the same irrespective of the specific improvement, either regular clover improvement (left) \citep{Sheikholeslami:1985ij} or exponential clover improvement (right) \citep{Francis:2019muy}. Due to the mathematical structure of the two improvements, the inverse of the regular clover improvement kernel takes an approximate factor of 2 longer than the kernel shown here for this local lattice of $48^4$. In the exponential clover, the diagonal part of the Dirac operator and its inverse take the same time to execute. This is particularly useful when using Hasenbusch acceleration \citep{Hasenbusch:2001ne}.

\section{Weak and strong scaling}\label{weak-and-strong-scaling}

We must stress that a good scaling up to 1000 GPUs only works for weak scaling. Here, we observe a close-to-optimal scaling. For a single GCD execution, we are achieving around 82\% (1.34 TB/s) of the theoretical peak bandwidth of the MI250X GCD, given by 1.6384 TB/s. For larger jobs, we are achieving less because of the overhead caused by separately calling the boundary kernels and the buffer synchronization. If the inner Dirac kernel execution is becoming too fast to mask the communications, additional overhead is expected causing an issue with strong scaling.

\begin{figure}

{\centering \includegraphics{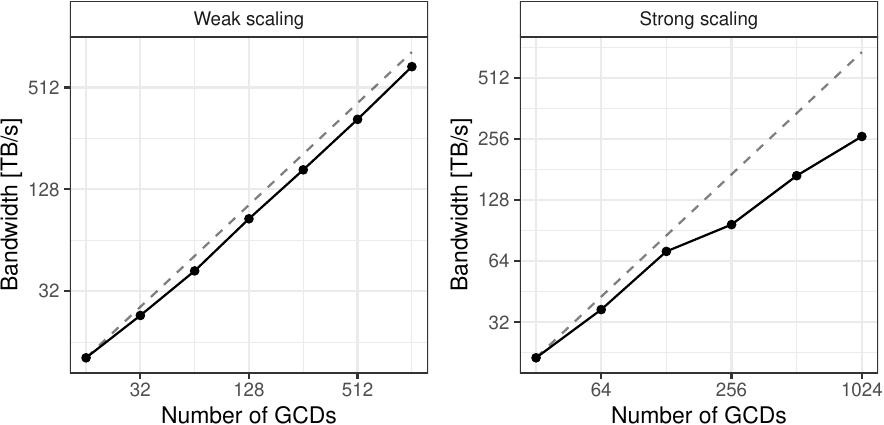} 

}

\caption{Weak and strong scaling from 16 to 1024 GCDs on LUMI-G with all four dimensions parallelized on a local lattice of $48^4$ (weak scaling) and global lattice of $128^4$ (strong scaling).}\label{fig:ws-ss}
\end{figure}

\begin{table}
\centering
\caption{\label{tab:ws-ss-table}Strong and weak scaling data for the unimproved Wilson-Dirac operator in terms of bandwidth reached on LUMI-G using AMD MI250X GCDs}
\centering
\begin{tabular}[t]{rrrrrrl}
\toprule
$T/a$ & $L_1/a$ & $L_2/a$ & $L_3/a$ & GCDs & Bandw. [TB/s] & [\%]\\
\midrule
96 & 96 & 96 & 96 & 16 & 12.93176 & 49\%\\
192 & 96 & 96 & 96 & 32 & 23.00843 & 44\%\\
384 & 96 & 96 & 96 & 64 & 42.18804 & 40\%\\
768 & 96 & 96 & 96 & 128 & 85.74323 & 41\%\\
1536 & 96 & 96 & 96 & 256 & 166.92260 & 40\%\\
\addlinespace
3072 & 96 & 96 & 96 & 512 & 331.60242 & 40\%\\
6144 & 96 & 96 & 96 & 1024 & 679.34508 & 40\%\\
128 & 128 & 128 & 128 & 32 & 21.38658 & 41\%\\
128 & 128 & 128 & 128 & 64 & 36.94729 & 35\%\\
128 & 128 & 128 & 128 & 128 & 71.44381 & 34\%\\
\addlinespace
128 & 128 & 128 & 128 & 256 & 96.84180 & 23\%\\
128 & 128 & 128 & 128 & 512 & 168.57047 & 20\%\\
128 & 128 & 128 & 128 & 1024 & 262.96682 & 16\%\\
\bottomrule
\end{tabular}
\end{table}

Fig.~\ref{fig:ws-ss} shows the scaling behavior on LUMI-G, with corresponding data in Tab.\ref{tab:ws-ss-table}. While the weak scaling is near perfect, achieving a peak bandwidth of 679 TB/s, jobs that contain over 128 GPUs are showing decreasing efficiency when scaling strongly. This is likely because the inner Dirac computations are becoming too fast to mask the communications even on a state-of-the-art network. While this is an issue for the efficiency of jobs, it is still possible to scale to relatively small local lattices and scale jobs to achieve unparalleled execution times. So far, for SU(2) gauge groups, we observe that local lattices smaller than \(16^4\) should be avoided, but for larger numbers of colors, this needs to be tested individually.

\section{Performance reached in comparison}\label{performance-reached-in-comparison}

\begin{figure}

{\centering \includegraphics{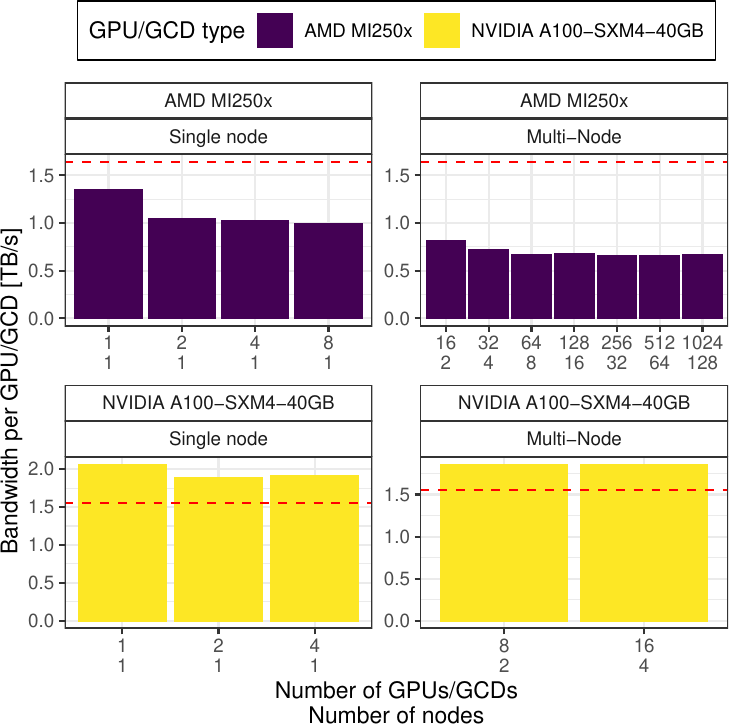} 

}

\caption{Comparison of bandwidths per GPU (Tursa) or GCD (LUMI-G) reached for single and multi-node jobs. The red dashed line denotes the peak theoretical bandwidth of the respective GPU/GCD.}\label{fig:comparison}
\end{figure}

In Fig. \ref{fig:comparison}, we compare the peak bandwidths reached on two different architectures. While we exceed the peak theoretical bandwidth on the NVIDIA-A100 on Tursa due to L2 reuse, we have yet to reach peak performance on the MI250X on LUMI-G. The impact of the overhead is also heavier in the case of LUMI-G. However, the scaling of the code for multi-node calculations is close to perfect and reaches 40\% of the peak performance of 1024 GCDs on LUMI-G.

\section{Conclusion and outlook}\label{conclusion-and-outlook}

We can reach 82\% (1.34 TB/s) of the peak performance on AMD MI250X cards for a code originally developed for CUDA. The code scales up to 1024 GCDs, reaching 679 TB/s. This increases the possibilities for the BSM physics lattice community to reach high precisions for predictions from the lattice.

\section{Acknowledgements}\label{acknowledgements}

This project has received funding from the European Union's Horizon 2020 research and innovation program under the Marie Sk\l odowska-Curie grant agreement \textnumero 813942. Testing, development, and benchmarking of this software was possible using resources on LUMI-G provided by the Danish eInfrastructure Consortium under grant application number DeiC-SDU-N5-2024055 and NVIDIA V100, A100, and H100 nodes provided by the UCloud DeiC Interactive HPC system managed by the eScience Center at the University of Southern Denmark. This work used the DiRAC Extreme Scaling service Tursa at the University of Edinburgh, managed by the Edinburgh Parallel Computing Centre on behalf of the STFC DiRAC HPC Facility (www.dirac.ac.uk). The DiRAC service at Edinburgh was funded by BEIS, UKRI and STFC capital funding and STFC operations grants. DiRAC is part of the UKRI Digital Research Infrastructure.

\renewcommand\refname{References}
  \bibliography{literature.bib}

\end{document}